\documentclass[12pt]{article}

\usepackage{amsmath}
\usepackage{amssymb}

\newcommand{\be}{\begin{equation}}
\newcommand{\ee}{\end{equation}}
\newcommand{\bea}{\begin{eqnarray*}}
\newcommand{\eea}{\end{eqnarray*}}
\newcommand{\bdm}{\begin{displaymath}}
\newcommand{\edm}{\end{displaymath}}

\newcommand{\Cal}[1]{\ensuremath{\mathcal{#1}}}

\newcommand{\calD}{\ensuremath{\mathcal{D}}}
\newcommand{\avg}[1]{\ensuremath{\langle{#1}\rangle}}

\newcommand{\aD}{\ensuremath{a_\calD}}
\newcommand{\aDdot}{\ensuremath{\dot a_\calD}}
\newcommand{\aDddot}{\ensuremath{\ddot a_\calD}}

\newcommand{\eqn}[1]{Eqn. \eqref{#1}}
\newcommand{\eqns}[1]{Eqns. \eqref{#1}}

\newcommand{\ph}[1]{\phantom{#1}}

\newcommand{\stavg}[1]{\ensuremath{\left\langle #1\right\rangle_{ST}}}
\newcommand{\avgD}[1]{\ensuremath{\left\langle #1\right\rangle_{\Cal{D}}}}

\newcommand{\om}{{\bf{\omega}}} 
\newcommand{\Om}[2]{{\bf{\Omega}{}^{#1}_{\ph{#1}#2}}} 
\newcommand{\barOm}[2]{\bar{\bf{\Omega}}{}^{#1}_{\ph{#1}#2}} 
\newcommand{\rmb}[1]{{\bf #1}} 
\newcommand{\ext}{{\rmb{d}}} 
\newcommand{\dx}[1]{\ext x^{#1}} 

\newcommand{\bil}[1]{\widetilde{#1}}

\newcommand{\Wxx}[2]{\ensuremath{\Cal{W}{}^{#1^\prime}_{#2}(x^\prime,x)}} 
\newcommand{\Wxxinv}[2]{\ensuremath{\Cal{W}{}^{#1}_{#2^\prime}(x,x^\prime)}}  
\newcommand{\W}[2]{\ensuremath{\Cal{W}{}^{#1^\prime}_{#2}}}

\newcommand{\Linh}{\ensuremath{L_{\rm inhom}}}
\newcommand{\Lfrw}{\ensuremath{L_{\rm FLRW}}}
\newcommand{\Lhub}{\ensuremath{L_{\rm Hubble}}}
\newcommand{\bt}{\ensuremath{\bar t}}
\newcommand{\ba}{\ensuremath{\bar a}}
\newcommand{\Mbar}{\ensuremath{\bar{\Cal{M}}}}

\newcommand{\uD}[1]{\ensuremath{#1_\Cal{D}}}

\newcommand{\bZ}[4]{{\rmb{Z}{}^{#1\ph{i}#3}_{\ph{i}#2\ph{i}#4}}} 

\begin{document}

\ \vskip 0.5 in

\begin{center}
 { \large {\bf The Effect of Cosmic Inhomogeneities On  }}

\smallskip

{\large  {\bf   { The Average Cosmological Dynamics} }}

\vskip 0.2 in

\smallskip

\bigskip

\bigskip

\bigskip

{\large{\bf T. P.  Singh}}

\medskip

{\it Tata Institute of Fundamental Research,}\\
{\it Homi Bhabha Road, Mumbai 400 005, India.}\\
{\tt email: tpsingh@tifr.res.in}\\
\medskip

\vskip 0.5cm
\end{center}

\vskip 1.0 in

\begin{abstract}

\noindent   It is generally assumed that on sufficiently large scales the Universe is well-described as a homogeneous, isotropic FRW cosmology with a dark energy.  Does the formation of nonlinear cosmic inhomogeneities produce a significant effect on the average large-scale FLRW dynamics? As an answer, we suggest that if the length  scale at which homogeneity sets in  is much smaller than the Hubble length scale, the back-reaction due to averaging over inhomogeneities is negligible.  This result is supported by more than one approach to study of averaging in cosmology. Even if no single approach is sufficiently rigorous and compelling, they are all in agreement that the effect of averaging  in the real Universe is small. On the other hand, it is perhaps fair to say that there is no definitive observational evidence yet that there indeed is a homogeneity scale which is much smaller than the Hubble scale, or for that matter, if today's Universe is indeed homogeneous on large scales. If the Copernican principle can be observationally established to hold, or is theoretically assumed to be valid, this provides strong evidence for homogeneity on large scales. However, even this by itself does not say what the scale of homogeneity is. If that scale is today comparable to the Hubble radius, only a fully non-perturbative analysis can establish or rule out the importance of cosmological back-reaction.  This brief elementary report summarizes some recent theoretical developments on which the above inferences are based.

\end{abstract}

\bigskip

\bigskip

\bigskip

\centerline{\it Based on a talk given at `International Conference on Two Cosmological Models'}
\centerline{17-19 November, 2010, Mexico City, To appear in the Conference Proceedings}

\newpage

\section{Introduction}
\noindent The Universe that we see around us is lumpy - it has stars, galaxies, clusters of galaxies, superclusters, sheets, filaments and voids. We do not precisely know from observations  what the size of the largest structures is; the  size beyond which there are no larger structures. On the other hand the early Universe is very well-described as a homogeneous, isotropic FRW  cosmology [Big-Bang nucleosynthesis and the relic CMB are evidence of success] and the present Universe is well-described as an FRW cosmology with dark energy. How does one reconcile a universe which is observed to be inhomogeneous and anisotropic on smaller scales, with a universe that is assumed to be homogeneous and isotropic on large scales? Clearly, some way of averaging the matter distribution and the related Einstein equations has to be invoked. What is the right way? The true metric of the universe is the one produced by the inhomogeneous matter distribution. On large scales, one assumes the average of the true metric to be FLRW, constructs the Einstein tensor for it, and uses it on the left hand side of the Friedmann equations wherein the matter content on the right hand side is a perfect fluid.  Because Einstein equations are nonlinear, the Einstein tensor constructed from the average metric tensor will in general not be the same as the average of the Einstein tensor of the true metric :
\begin{equation}
<G_{\mu\nu} (g_{\mu\nu})>\  = \ <T_{\mu\nu}> \ \neq \ G_{\mu\nu} (<g_{\mu\nu}>)
\label{av}
\end{equation}
Here $<...>$ denotes the averaging operation [whose correct definition, for tensors on a curved spacetime, is itself a major challenge]; and $<g_{\mu\nu}>=g_{\mu\nu}|_{FLRW}$. The correct  averaged Einstein equations are of course given by the first pair (the equality) in the above set, whereas in cosmology we  assume the correct equations to be those given by the second pair (by assuming the inequality to actually be an equality).  This is obviously done because the latter option is infinitely simpler - it is straightforward to write the Einstein tensor for the Robertson-Walker metric, but it is impossible to find the true metric of the inhomogeneous universe and then average its corresponding Einstein tensor. Since the first and the third terms in (\ref{av}) could differ significantly, we might be working with the wrong averaged Einstein equation on cosmological scales. This is the averaging problem: how to correctly average Einstein equations, and to find out if the neglected terms (the so-called back-reaction) can become important in the late stages of an evolving universe, when nonlinear structures such as galaxies and clusters form. In particular, can the back-reaction mimic a dark energy, and explain the observed cosmic acceleration?

The problem of  averaging of Einstein equations has a long history, and has recently been reviewed in an important article by Ellis \cite{Ellis}. Important contributions to the study of averaging have been made in recent times, amongst others, by Buchert \cite{Buchert}, Coley \cite{Coley}, Wald \cite{Wald}, Zalaletdinov \cite{Zalaletdinov} and their collaborators. Specific applications have been developed by Kolb \cite{Kolb}, Marra \cite{Marra}, Rasanen \cite{Rasanen}, Sussman \cite{Sussman}, Wiltshire \cite{Wiltshire} and others. Much of this work, as well as earlier developments, are reviewed by Ellis, and we will not enter into details here, except in the context of specific arguments developed here. 

If we want to find out the back-reaction on an FLRW universe, it certainly means we are taking an FLRW geometry as given on large scales. It is hence necessary to first know what the observational evidence for large scale homogeneity and isotropy is, and what is the length scale at which homogeneity sets in. 

\section{Evidence for large scale homogeneity}

Careful discussions of this issue have recently been given by Clarkson and Maartens \cite{CM}, Maartens \cite{Maartens} and Ellis \cite{Ellis}. What we have to say below is a summary from these earler works, and is reported here because of its significance for the discussion on averaging in the next section. 

It is well known that homogeneity on spatial surfaces can not be established by direct observations, because all our observations are on the past light-cone. Hence tests of homogeneity have to exploit the following route : isotropy around us along with the Copernican Principle [CP] implies homogeneity, and hence a FLRW Universe.  So one needs to test for isotropy and for CP independently.

The observational evidence for spacetime isotropy around our world-line can be investigated from examining the isotropy of the CMB and the galaxy distribution. For a perfectly  isotropic CMB, all multipoles of the distribution function higher than the monopole, as well their time derivatives, vanish. However without CP one cannot deduce the vanishing of the spatial derivatives of the higher multipoles, and hence spacetime isotropy about our world-line cannot be deduced without CP. As for baryonic matter (along with certain assumptions for the distribution of CDM and dark energy) it can be shown that isotropic distribution of the following four matter observables  on the light-cone implies an isotropic spacetime geometry : angular diameter distances, galaxy number counts, bulk velocities and lensing (details and references to original work can be found in \cite{Maartens}). As pointed out by Maartens, it is not known whether almost-isotropy of observations leads to almost-isotropy of spacetime geometry.

Next, one considers what can be inferred about spatial homogeneity, if one assumes CP, and considers the following three cases : isotropic matter distribution, isotropic CMB, almost-isotropic CMB. If all fundamental observers measure the same isotropic distribution of the four matter observables mentioned above, this implies homogeneity, and the Universe is FLRW. It can be proved that exact isotropy of the CMB for all observers also implies an FLRW universe. Almost-isotropy of the CMB can be shown, via a 
non-perturbative analysis, to imply an almost-FLRW universe, provided some of the time and spatial derivatives of the multipoles are sufficiently small.

Thus it is clear that the case for an almost-FLRW universe will be strong if observational tests support the Copernican Principle. These tests can be carried out by testing the standard consistency relations in FLRW geometry. The FLRW curvature parameter which can be inferred from geometric measurements is independent of redshift, and a detection of redshift dependence of this parameter will indicate departure from homogeneity. A second test is the time drift of cosmological redshift, and a third test is a significant difference between the radial and transverse BAO scales. None of these tests have yet been carried out, but their eventual execution will play a crucial role in confirming large-scale homogeneity. The CP can also be tested by looking for a large thermal or kinetic Sunyaev-Zeldovich effect temeperature distortion of the CMB. Also, a large SZ effect induced CMB polarization could indicate a violation of CP and hence of homogeneity.

As of now, there is no evidence against CP, but neither is there clinching evidence for large-scale homogeneity. Also, it is not quite clear at what scale homogeneity sets in. If we assume that there is homogeneity, and that too at a scale much less than the Hubble scale, say at around $100$ MPc, then it can be shown (as discussed next) that the cosmological back-reaction is negligible. And the $\Lambda$CDM model is then a good description of the present day Universe. On the other hand  if there are much larger nonlinear structures in the Universe - their formation can then no longer be described perturbatively on an FLRW background, and the back-reaction problem will have to be examined afresh. 

\section{Averaging in Cosmology and Calculation of Back-Reaction} 

Assuming that the scale at which homogeneity sets in is much smaller than the Hubble scale, we give three explanations as to why the back-reaction will be small : (i) a simple argument due to Peebles \cite{Peebles}; (ii) our own work \cite{AS1}, \cite{AS2}, \cite{A3}, \cite{AS4} which builds on Zalaletdinov's Macroscopic Theory of Gravity [MG] \cite{Z1}, \cite{Z2}, \cite{Z3}, \cite{Z4}; and (iii) the work  of Wald and collaborators \cite{W2}, \cite{Wald}. Similar results obtained by a few other researchers, which support the present inference, are briefly reviewed in Paranjape's thesis \cite{APthesis}. 

\subsection{An argument due to Peebles (\cite{Peebles} and references therein)} 

For nonlinear structures such as galaxies, the Newtonian gravitational potential is of the order of the square of the velocity dispersion [about $300$ km/sec], i.e. $\phi\sim 10^{-6}$. Hence the galaxy distribution can be described as a perturbation over an FLRW universe. The metric can be written as a perturbed FLRW Universe and the Einstein equations can be split into an evolution equation for the background scale-factor and the Poisson equation for the perturbed Newtonian potential determined by the density contrast (assumed to be provided by non-interacting dark matter).  

In order to find the effect of averaging on the FLRW equations, spatial averages of Einstein equations need to be computed to order $\phi^2$, in particular for the dominant term which is proportional to $\nabla\phi.\nabla\phi$. When this is done, one finds corrections due to back-reaction in  both the Friedmann equations - corrections in the form of a kinetic energy coming from the mean square velocity dispersion, and the averaged gravitational potential energy determined by the density contrast of the formed nonlinear structures. Both these correction terms are of the order of a part in a million, and hence much smaller than the magnitude of the observed dark energy.

The discussion by Peebles is patterned in part on the nice work of Siegel and Fry \cite{SF}. It seems to us that there is room for improvement in this argument : one should not fix the background, but allow for the possibility that as perturbations grow, the background around which back-reaction should be calculated may itself be changing, because of feedback from the perturbations. One has to ascertain that a runaway process leading to breakdown of perturbation theory does not take place. This is the study we attempted by applying Zalaletdinov's  averaging theory [Macroscopic Gravity] to cosmology. Before we summarize our work on applying MG,  it will be useful to briefly review Buchert's averaging scheme. We do this because the Buchert approach provides simple averaged equations, while being less ambitious than MG. Remarkably, the averaged equations that arise from MG are very similar to Buchert's averaging equations, enforcing a certain high degree of reliability of both approaches, despite their conceptual differences.

\subsection{Buchert's averaging scheme for a dust spacetime}

For a general spacetime containing irrotational dust, the metric can be written as
\begin{equation}
ds^2=\,-dt^2+h_{ij}(\vec{x},t)dx^idx^j\,.
\label{avg1}
\end{equation}
The expansion tensor $\Theta^i_j$ is given by $\Theta^i_j\equiv
(1/2)h^{ik}\dot h_{kj}$ where the dot refers to a derivative with
respect to  time $t$. The traceless symmetric shear tensor is defined
as  $\sigma^i_j\equiv \Theta^i_j-(\Theta/3) \delta^i_j$ where $\Theta
=\Theta^i_i$  is the expansion scalar. The Einstein equations can be
split into a set of scalar equations and a set of vector
and traceless tensor equations. The scalar equations are the
Hamiltonian constraint \eqref{avg2a} and the evolution equation for
$\Theta$ \eqref{avg2b},
\begin{subequations}
\begin{equation}
^{(3)}\mathcal{R}+\frac{2}{3}\Theta^2-2\sigma^2=16\pi G\rho
\label{avg2a}
\end{equation}
\begin{equation}
^{(3)}\mathcal{R}+\dot\Theta+\Theta^2=12\pi G\rho
\label{avg2b}
\end{equation}
\end{subequations}
where the dot denotes derivative with respect to time $t$,
$^{(3)}\mathcal{R}$ is the Ricci scalar of the 3-dimensional hypersurface
of  constant $t$ and $\sigma^2$ is the rate of shear  defined by
$\sigma^2\equiv (1/2)\sigma^i_j\sigma^j_i$. Eqns. \eqref{avg2a} and
\eqref{avg2b} can be combined to give Raychaudhuri's equation
\begin{equation}
\dot\Theta+\frac{1}{3}\Theta^2+2\sigma^2+4\pi G\rho=0\,.
\label{avg3}
\end{equation}
The continuity equation $\dot\rho=-\Theta\rho$ which gives the
evolution of $\rho$, is consistent with Eqns. \eqref{avg2a},
\eqref{avg2b}. We only consider the scalar equations, since the
spatial average of a   scalar quantity can be defined in a gauge
covariant manner within a given  foliation of space-time.
For the space-time described by \eqref{avg1}, the spatial average of a scalar
$\Psi(t,\vec{x})$ over a {\em comoving}  domain \calD\ at time $t$ is
defined by
\begin{equation}
\avg{\Psi}=\frac{1}{V_\calD}\int_\calD{d^3x\sqrt{h}\,\Psi}
\label{avg4}
\end{equation}
where $h$ is the determinant of the 3-metric $h_{ij}$ and $V_\calD$ is
the volume of the comoving domain given by
$V_\calD=\int_\calD{d^3x\sqrt{h}}$.

Spatial averaging is, by definition, not generally covariant. Thus the
choice of foliation is relevant, and should be motivated on physical
grounds. In the context of cosmology, averaging over freely-falling
observers is a natural choice, especially when one intends to compare
the results with standard FRW cosmology. Following the definition
(\ref{avg4}) the following commutation relation then holds
\cite{Buchert}  
\begin{equation}
\avg{\Psi}^\cdot-\avg{\dot\Psi}=
\avg{\Psi\Theta}-\avg{\Psi}\avg{\Theta}
\label{avg5}
\end{equation}
which yields for the expansion scalar $\Theta$
\begin{equation}
\avg{\Theta}^\cdot-\avg{\dot\Theta}=
\avg{\Theta^2}-\avg{\Theta}^2\,.
\label{avg6}
\end{equation}
Introducing the dimensionless scale factor
$\aD\equiv\left(V_\calD/V_{\calD in}\right)^{1/3}$ normalized by the
volume of the domain \calD\ at some initial time $t_{in}$, we can
average the scalar Einstein equations \eqref{avg2a}, \eqref{avg2b} and
the continuity  equation to obtain
\begin{subequations}
\begin{equation}
\avg{\Theta}=3\frac{\aDdot}{\aD}\,,
\label{avg7a}
\end{equation}
\begin{equation}
3\left(\frac{\aDdot}{\aD}\right)^2-8\pi G\avg{\rho}+
\frac{1}{2}\avg{\mathcal{R}}=\,-\frac{\mathcal{Q}_\calD}{2}\,,
\label{avg7b}
\end{equation}
\begin{equation}
3\left(\frac{\aDddot}{\aD}\right)+4\pi
G\avg{\rho}=\mathcal{Q}_\calD\,,
\label{avg7c}
\end{equation}
\begin{equation}
\avg{\rho}^\cdot=\,-\avg{\Theta}\avg{\rho}=\,-
3\frac{\aDdot}{\aD}\avg{\rho}\,.
\label{avg7d}
\end{equation}
\end{subequations}
Here $\avg{\mathcal{R}}$, the average of the spatial Ricci scalar
$^{(3)}\mathcal{R}$, is a domain dependent spatial constant. The
`backreaction' $\mathcal{Q}_\calD$ is given by
\begin{equation}
\mathcal{Q}_\calD\equiv\frac{2}{3}\left(\avg{\Theta^2}-
\avg{\Theta}^2\right)-2\avg{\sigma^2}
\label{avg8}
\end{equation}
and is also a spatial constant. The last equation \eqref{avg7d} simply
reflects the fact that the mass contained in a comoving domain is
constant by construction : the local continuity equation
$\dot\rho=-\Theta\rho$ can be solved to give
$\rho\sqrt{h}=\rho_0\sqrt{h_0}$ where the subscript $0$ refers to some
arbitrary reference time $t_0$. The mass $M_\calD$ contained in a
comoving domain \calD\ is then $M_\calD=\int_\calD{\rho\sqrt{h}d^3x}
=\int_\calD{\rho_0\sqrt{h_0}d^3x}=\,$constant. Hence
\begin{equation}
\avg{\rho}=\frac{M_\calD}{V_{\calD in} \aD^3}
\label{avg9}
\end{equation}
which is precisely what is implied by Eqn. \eqref{avg7d}.

This averaging procedure can only be applied for spatial scalars, and
hence only a subset of the Einstein equations can be smoothed out. As
a result it may appear that the outcome of such an approach is
severely restricted, and essentially incomplete due to the
impossibility to analyze the full set of equations. However one should
note that the cosmological parameters of interest are scalars, and the
averaging of the exact scalar part of Einstein equations provides the
requisite needed information. A more general strategy would be to
consider the smoothing of tensors, which is beyond the scalar approach
that certainly provides useful information, albeit not the full
information.

The dynamical equations above can be cast in a form which is
immediately comparable with the standard FRW equations
\cite{Buchert}. Namely,
\begin{subequations}
\begin{equation}
\frac{\aDddot}{\aD}=\,-\frac{4\pi G}{3}\left(\rho_{\rm eff}+
3P_{\rm eff}\right)
\label{avg10a}
\end{equation}
\begin{equation}
\left(\frac{\aDdot}{\aD}\right)^2=\frac{8\pi G}{3}\rho_{\rm eff}
\label{avg10b}
\end{equation}
\end{subequations}
with $\rho_{\rm eff}$ and $P_{\rm eff}$ defined as
\begin{equation}
\rho_{\rm eff}=\avg{\rho}-\frac{\mathcal{Q}_\calD}{16\pi G}-
\frac{\avg{\mathcal{R}}}{16\pi G}~~~\text{;}~~~
P_{\rm eff}=\,-\frac{\mathcal{Q}_\calD}{16\pi
  G}+\frac{\avg{\mathcal{R}}}{48\pi G}\,.
\label{avg111}
\end{equation}
A necessary condition for \eqref{avg10a} to integrate to
\eqref{avg10b} takes the form of the following differential equation
involving $\mathcal{Q}_\calD$ and $\avg{\mathcal{R}}$
\begin{equation}
\mathcal{\dot Q}_\calD+6\frac{\aDdot}{\aD}\mathcal{Q}_\calD
+\avg{\mathcal{R}}^{\cdot}+2\frac{\aDdot}{\aD}\avg{\mathcal{R}}=0
\label{avg12}
\end{equation}
and the criterion to be met in order for the effective scale factor
$\aD$ to accelerate, is
\begin{equation}
\mathcal{Q}_\calD>4\pi G\avg{\rho}\,.
\label{avg13}
\end{equation}

The Buchert scheme has been applied extensively, and in particular can be used to show that there indeed are toy cosmological models which when averaged over inhomogeneities can produce  an apparent acceleration. However, not all Einstein equations are averaged, and one does not have an averaged metric here [which we would like to be the FLRW metric]. Macroscopic Gravity can achieve that, while reproducing modified Friedmann equtions analogous to the Buchert equations, when applied to cosmology.

\section{Macroscopic Gravity}

This theory is developed comprehensively in the works of Zalaletdinov; briefly introduced in \cite{AS1}, and reviewed in Paranjape's thesis \cite{APthesis}.  For detailed discussions and the primary interpretation of MG, the reader is referred to Zalaletdinov's original papers cited in this article.  

For the purpose of averaging of tensors the key new element which is introduced is a \emph{bivector} \Wxx{a}{b}\  which transforms as a vector at event $x^\prime$ and as a co-vector at event $x$. The bivector is used to define the ``bilocal 
extension'' of a general tensorial object  
\begin{equation}
\bil{P}{}^a(x^\prime,x) =
\Wxxinv{a}{a}P{}^{a^\prime}(x^\prime) \,
\label{avgZala3}
\end{equation}
The ``average'' of $P{}^a(x)$ over a 4-dimensional spacetime region $\bf{\Sigma}$ with a supporting point $x$ is
\begin{equation}
\bar P{}^a(x) = \stavg{\bil{P}{}^a} = \frac{1}{V_\Sigma}
\int_\Sigma{d^4x^\prime\sqrt{-g^\prime}\bil{P}{}^a(x^\prime,x)} \,
 \label{avgZala4}
\end{equation}
and this averaging operation preserves tensorial properties.

There is a certain degree of non-uniqueness in te choice of the coordination bi-vector - the freedom coming from the presence of undetermined structure constants in the commutation relations for a vector basis in terms of which one can solve for the cordination bivector. The simplest choice is to set these structure constants to zero. When that is done, then
in a volume preserving coordinate system $\phi^m$, [VPC], i.e. one with $g(\phi^m)=$\ constant, the coordination bivector takes its most simple form, namely 
\begin{equation}
\Wxx{a}{j}\mid_{\rm proper} = \delta{}^{a^\prime}_j\,.
\label{avgZala27}
\end{equation}
The effect of this non-uniqueness on the physical results for averaging in cosmology remains to be estimated. Nonetheless, it is noteworthy that the averaged Friedmann equations to be derived from this approach are similar to Buchert's and the physical results about the magnitude of the back-reaction is identical to the one due to Peebles. This gives confidence in the robustness of the results obtained, even though there is freedom in the choice of the coordination bivector.   It is also useful to note that this bi-vector is different from the Synge bi-tensor which leaves the metric invariant upon avaeraging, and hence cannot really be used to average an inhomogeneous geometry.  

Averaged Geometry : the key idea of Macroscopic Gravity is that the average connection $\barOm{a}{b}(x)$ 
\be
\barOm{a}{b} \equiv \avg{\Om{a}{b}}\,,
\label{MG-avgcond}
\ee
is defined as the connection $1$-form on a new, averaged manifold \Mbar. Next one defines a correlation $2$-form   
\begin{equation}
\bZ{a}{b}{i}{j} = \stavg{\Om{a}{b}\wedge\Om{i}{j}} -
\barOm{a}{b}\wedge\barOm{i}{j}\,.
\label{avgZala38}
\end{equation} 
Denoting
$\rmb{R}^a_{\ph{a}b}\equiv\stavg{\,\bil{\rmb{r}}^a_{\ph{a}b}}$ , where $\rmb{r}^{a}_{ b}$ is the curvature 2-form of the inhomogeneous geometry, and the
curvature $2$-form on the averaged manifold \Mbar\ as
$\rmb{M}^a_{\ph{a}b}$ can be shown to give
\begin{equation}
\rmb{M}^a_{\ph{a}b} = \rmb{R}^a_{\ph{a}b} - \bZ{a}{c}{c}{b}\,.
\label{avgZala40b}
\end{equation}
The inhomogeneous Einstein equations 
\begin{equation}
g^{ak}r_{kb} - \frac{1}{2}\delta{}^a_b g^{ij}r_{ij} = -\kappa
t{}^{a({\rm mic})}_b\,,
\label{avgZala56}
\end{equation}
average out to
\begin{equation}
E{}^a_b = -\kappa T{}^a_b + C{}^a_b\,,
\label{avgZala61}
\end{equation}

\begin{equation}
C{}^a_b = \left(Z^a_{\ph{a}ijb} - \frac{1}{2}\delta{}^a_b
Z^m_{\ph{a}ijm} \right)G^{ij}.
\label{correln10}
\end{equation}
$G^{ij}$ is the metric on the averaged geometry.
The correlation 2-form is assumed to satisfy certain differential conditions which amount to closure conditions for the above system of averaged equations. 

These averaged equations of Macroscopic Gravity carry, in a covariant and non-perturbative manner, information about the effect of the underlying inhomogeneities on the averaged geometry. 

\section{Application of Macroscopic Gravity to Cosmology}

In order to apply MG to cosmology we start with the assumption that
Einstein's equations are to be imposed on length scales where
stars are pointlike objects (we denote such a scale as \Linh). The
averaging we perform will be directly at a length scale \Lfrw\
larger than about $100h^{-1}$Mpc or so. This averaging scale is
assumed to satisfy $\Linh\ll\Lfrw\ll\Lhub$ where \Lhub\ is the
length scale of the observable universe. The averaging will be
assumed to yield a geometry which has homogeneous and isotropic
spatial sections. In other words, we will assume that the averaged
manifold \Mbar\ admits a preferred, hypersurface-orthogonal unit
timelike vector field $\bar v^a$, which defines $3$-dimensional
spacelike hypersurfaces of constant curvature, and that $\bar v^a$
is tangent to the trajectories of observers who see an isotropic
Cosmic Background Radiation. (These ``observers'' are defined in
the averaged manifold -- we will clarify below what they
correspond to in the inhomogeneous manifold.) Throughout the rest
of this article, for simplicity, we will work with the special case
where the spatial sections on \Mbar\ defined by $\bar v^a$ are
flat. (In principle the entire calculation can be repeated for
non-flat spatial sections as well.) One can then choose
coordinates $(t,x^A)$, $A=1,2,3$, on \Mbar\ such that the spatial
line element takes the form
\begin{equation}
^{(\Mbar)}ds^2_{\rm spatial} = a^2(t)\delta_{AB}dx^Adx^B\,,
\label{spatlim1}
\end{equation}
where $\delta_{AB}=1$ for $A=B$, and $0$ otherwise, and we have
$\bar v^a=(\bar v^t,0,0,0)$ so that the spatial coordinates are
comoving with the preferred observers. The vector field $\bar v^a$
also defines a proper time (the cosmic time) $\tau$ such that
$\partial_\tau = \bar v^a\partial_a = \bar v^t\partial_t$. We will
further assume that the averaged energy-momentum tensor $T{}^a_b$
can be written in the form of a perfect fluid, as
\begin{equation}
T{}^a_b = \rho \bar v^a \bar v_b + p\pi^a_b\,, \label{spatlim2}
\end{equation}
where the projection operator $\pi^a_b$ is defined as
\begin{equation}
\pi^a_b = \delta{}^a_b + \bar v^a\bar v_b \,, \label{spatlim3}
\end{equation}
and $\rho$ and $p$ are the homogeneous energy density and pressure
respectively, as measured by observers moving on trajectories (in
\Mbar) with the tangent vector field $\bar v^a$,
\begin{equation}
\rho\equiv T{}^a_b\bar v^b\bar v_a ~~;~~ p\equiv\frac{1}{3}\pi^b_a
T{}^a_b\,. \label{spatlim-T-ab1}
\end{equation}
$\rho$ and $p$ are observationally relevant quantities, since all
measurements of the matter energy density, especially those from
studies of Large Scale Structure, interpret observations in the
context of the averaged geometry.  An important consequence of the above assumptions is that
the correlation tensor $C{}^a_b$, when expressed in terms of the
natural coordinates adapted to the spatial sections defined by the
vector field $\bar v^a$, is \emph{spatially homogeneous}. This is
clear when the modified Einstein equations  are
written in these natural coordinates.

The existence of the vector field $\bar v^a$ with the attendant
assumptions described above, allows us to separate out the
nontrivial components of the (FLRW) Einstein tensor $E{}^a_b$ on
\Mbar\ in a coordinate independent fashion -- the Einstein tensor
can be written as
\begin{align}
E{}^a_b = j_1(x) \bar v^a\bar v_b &+ j_2(x) \pi^a_b \nonumber\\
j_1(x) \equiv E{}^a_b\bar v^b\bar v_a ~~&;~~ j_2(x) \equiv
\frac{1}{3}\left( \pi^b_a E{}^a_b\right)\,, \label{spatlim-FLRW1}
\end{align}
where $j_1(x)$ and $j_2(x)$ are scalar functions whose form
depends upon the coordinates used. The remaining components given
by $\pi^b_k E{}^a_b\bar v_a$ and the traceless part of
$\pi^i_a\pi^b_k E{}^a_b$, vanish identically. Since the
energy-momentum tensor $T{}^a_b$ in \eqn{spatlim2} also has an
identical structure, this structure is therefore \emph{also
imposed} on the correlation tensor $C{}^a_b$. Namely, $\pi^b_k
C{}^a_b\bar v_a$ and the traceless part of $\pi^i_a\pi^b_k
C{}^a_b$ \emph{must vanish}. This is a condition on the underlying
inhomogeneous geometry, irrespective of the coordinates used on
either \Cal{M}\ or \Mbar, and is clearly a consequence of
demanding that the averaged geometry have the symmetries of the
FLRW spacetime. 

This leads us to the crucial question of the choice of
\emph{gauge} for the underlying geometry : namely, what choice of
spatial sections for the \emph{inhomogeneous} geometry, will lead
to the spatial sections of the FLRW metric in the comoving
coordinates defined in \eqn{spatlim1}? Since the matter
distribution at scale \Linh\ need not be pressure-free (or,
indeed, even of the perfect fluid form), there is clearly no
natural choice of gauge available, although locally, a synchronous
reference frame can always be chosen. We note that there must be
\emph{at least one} choice of gauge in which the averaged metric
has spatial sections in the form \eqref{spatlim1} -- this is
simply a refinement of the Cosmological Principle, and of the Weyl
postulate, according to which the Universe is homogeneous and
isotropic on large scales, and individual galaxies are considered
as the ``observers'' travelling on trajectories with tangent $\bar
v^a$. In the averaging approach, it makes more sense to replace
``individual galaxies'' with the \emph{averaging domains}
considered as physically infinitesimal cells -- the ``points'' of
the averaged manifold \Mbar. This is physically reasonable since
we know after all, that individual galaxies exhibit peculiar
motions, undergo mergers and so on. This idea is also more in
keeping with the notion that the Universe is homogeneous and
isotropic \emph{only on the largest
  scales}, which are much larger than the scale of individual
galaxies.

Consider any $3+1$ spacetime splitting in the form of a lapse
function ${N}(t,x^J)$, a shift vector ${N}^A(t,x^J)$, and a metric
for the $3$-geometry ${h}_{AB}(t,x^J)$, so that the line element
on \Cal{M}\ can be written as
\begin{equation}
^{(\Cal{M})}ds^2 = -\left({N}^2 -
  {N}_{\!A}{N}^A\right)dt^2 + 2{N}_{\!B}dx^Bdt +
  {h}_{AB}dx^Adx^B\,,
\label{spatlim4}
\end{equation}
where ${N}_{\!A} = {h}_{\!AB}{N}^B$. At first sight, it might seem
reasonable to leave the choice of gauge arbitrary. However the analysis is then complicated. On the other hand, if we make the assumption that
the spatial sections on \Cal{M}\ leading to the spatial metric
\eqref{spatlim1} on \Mbar, are spatial sections \emph{in a volume
  preserving gauge}, then the correlation terms simplify
greatly. This is not surprising since the MG formalism is nicely
adapted to the choice of volume preserving coordinates. The case when the gauge is left unspecified is dealt with in our original papers.

To begin our calculation, we perform a coordinate
transformation and shift to the gauge wherein the new lapse
function $N$ is given by $N=1/\sqrt{h}$ where $h$ is the
determinant of the new $3$-metric $h_{AB}$. In general, one will
now be left with a non-zero shift vector $N^A$; however, the
condition $N\sqrt{h}=1$ ensures that the coordinates we are now
using are volume preserving, since the metric determinant is given
by $g=-N^2h=-1={\rm constant}$. We denote these volume preserving
coordinates (VPCs) by $(\bt, \rmb{x}) = (\bt, x^A) = (\bt, x, y,
z)$, and will assume that the spatial coordinates are non-compact.
For simplicity, we make the added assumption that $N^A=0$ in the
inhomogenous geometry, so that
$g_{\bt\,\bt}=-N^2=-1/h$ and $g_{\bt A}=0$. The line element for
the inhomogenous manifold \Cal{M}\ becomes
\begin{equation}
^{(\Cal{M})}ds^2=-\frac{d\bt^2}{h(\bt,\rmb{x})} +
  h_{AB}(\bt,\rmb{x})dx^Adx^B\,.
\label{spatlim5}
\end{equation}
Note that in this gauge, the average takes on a particularly
simple form : for a tensor $p{}^i_j(x)$, with a spacetime
averaging domain given by the ``cuboid'' ${\Sigma}$ defined by
\begin{equation}
{\Sigma} =
\left\{(\bt,x,y,z)\mid-T/2<\bt<T/2,-L/2<x,y,z<L/2\right\},
\label{spatlim6}
\end{equation}
where $T$ and $L$ are averaging time and length scales
respectively, the average is given by
\begin{align}
&\stavg{\bil{p}{\,}^i_j}(\bt,\rmb{x}) =
    \stavg{p{}^i_j}(\bt,\rmb{x}) \nonumber\\ 
&=  \frac{1}{TL^3}
    \int_{\bt-T/2}^{\bt+T/2}{dt^\prime\int_{-L/2}^{+L/2}{
    dx^\prime dy^\prime dz^\prime\bigg[
      p{}^i_j(t^\prime,x^\prime,y^\prime,z^\prime)\bigg]}} \,,
\label{spatlim7}
\end{align}
where the limits on the spatial integral are understood to hold for
all three spatial coordinates. We define the ``spatial averaging
limit'' as the limit $T\to0$ (or $T\ll\Lhub$) which is interpreted as
providing a definition of the average on a spatial domain
corresponding to a ``thin'' time slice, the averaging operation now
being given by 
\begin{align}
&\avg{p{}^i_j}(\bt,\rmb{x}) \nonumber\\
&=  \frac{1}{L^3}
    \int_{-L/2}^{+L/2}{dx^\prime dy^\prime dz^\prime\bigg[ 
      p{}^i_j(\bt,x^\prime,y^\prime,z^\prime)\bigg]} +
    \Cal{O}\left(TL_{\rm Hubble}^{-1} \right) \,.
\label{spatlim8}
\end{align}
(Note the time dependence of the integrand.) Henceforth, averaging
will refer to spatial averaging, and will be denoted by
$\avg{...}$, in contrast to the spacetime averaging considered
thus far (denoted by $\stavg{...}$). The choice of a cube with
sides of length $L$ as the spatial averaging domain was arbitrary,
and is in fact not essential for any of the calculations to
follow. In particular, all calculations can be performed with a
spatial domain of arbitrary shape. We will only use
the cube for definiteness and simplicity in displaying equations.
The significance of introducing a
spatial averaging in this manner is that the construction of
spatial averaging is not isolated from spacetime averaging, but is
a special limiting case of the latter and is, in fact, still a
fully covariant operation.


For the volume preserving gauge, we have
\begin{align}
G_{\bt\bt} &= \avg{g_{\bt\,\bt}} = \avg{\frac{-1}{h}} = -f^2(\bt) ~;
\nonumber\\ 
&G_{AB} = \avg{h_{AB}}=\ba^2(\bt)\delta_{AB} \,,
\label{spatlim9}
\end{align}
where $\ba$ and $f$ are some functions of the time coordinate
alone. A few remarks are in order on this particular choice of
assumptions. Apart from the fact that the spacetime averaging
operation takes on its simplest possible form \eqref{spatlim7} in
this gauge and allows a transparent definition of the spatial
averaging limit, it can also be shown that the assumptions in
\eqn{spatlim9} are sufficient to establish the following relations
:
\begin{equation}
f^2(\bt) = \avg{\frac{1}{h}} = \frac{1}{\avg{h}} =
\frac{1}{\ba^6}\,. \label{spatlim10}
\end{equation}
Here the second equality arises from the condition $\bar
g^{ij}=G^{ij}$ which can be assumed whenever the averaged metric
is of the FLRW form. The last
equality follows on considering the conditions
$\avg{\bil{\Gamma}{}^a_{bc}} =\,^{(\rm FLRW)}\Gamma{}^a_{bc}$ in
obvious notation, (the basic assumption of the MG averaging
scheme). \eqn{spatlim10} reduces the line element on
\Mbar\ to the form
\begin{equation}
^{(\Mbar)}ds^2 = -\frac{d\bt^2}{\ba^6(\bt)} +
  \ba^2(\bt)\delta_{AB}dx^Adx^B \ .
\label{spatlim11}
\end{equation}
The line element in \eqn{spatlim11} clearly corresponds to the
FLRW metric in a \emph{volume preserving gauge}. In other words,
the (spatial) average of the inhomogeneous geometry in the volume
preserving gauge leads to a geometry with homogeneous and
isotropic spatial sections, also in a volume preserving gauge.
Note that the gauge in \eqn{spatlim11} for the FLRW spacetime
differs from the standard synchronous and comoving gauge, only by
a redefinition of the time coordinate. The vector field $\bar v^a$
introduced at the beginning of this section and which defines the
FLRW spatial sections, is now given by
\begin{equation}
\bar v^a = \left(\ba^3,0,0,0\right) ~~;~~ \bar v_a = G_{ab}\bar
v^b = \left(-\frac{1}{\ba^3},0,0,0,\right) \,. \label{spatlim12}
\end{equation}
%

Before proceeding to the calculation of the correlation terms and
the averaged Einstein equations, we briefly describe why it is
important to consider the spatial averaging limit of the MG
averaging operation. The key idea to emphasize is that an average
of the homogeneous and isotropic FLRW geometry, should give back
the same geometry. Since the FLRW geometry has a preferred set of
spatial sections, it is important therefore to perform the
averaging over these sections. Further, since the FLRW metric
adapted to its preferred spatial sections depends on the time
coordinate, it is also essential that the spacetime average should
involve a time range that is short compared to the scale over
which say the scale factor changes significantly. Clearly then, averaging the FLRW metric (denoted
$^{(FLRW)}g_{ab}$) given in \eqn{spatlim11} (which is in volume
preserving gauge) will strictly yield the same metric \emph{only} in
the limit $T\to0$. Namely, for the cuboid ${\Sigma}$ defined in
\eqn{spatlim6} 
\begin{align}
\avg{^{(FLRW)}\bil{g}_{ab}} &=
\lim_{T\to0}\frac{1}{TL^3}\int_\Sigma{dt^\prime d^3x^\prime\,
  ^{(FLRW)}g_{ab}(t^\prime,\rmb{x}^\prime)} \nonumber\\ 
&=\,^{(FLRW)}g_{ab}\,, \label{spatlim14}
\end{align}
which should be clear from the definition of the metric. The
result $\avg{^{(FLRW)}\bil{g}_{ab}} = \,^{(FLRW)}g_{ab}$ in the
spatial averaging limit can also be shown to hold for the FLRW
metric in synchronous gauge, where the coordination bivector
$\W{a}{j}$ can be easily computed using the transformation from
the VPCs $(\bt,x^A)$ to the synchronous coordinates $(\tau,y^A)$
given by
\begin{equation}
\tau = \int^{\bt}{\frac{dt}{\ba^3(t)}} ~~;~~ y^A = x^A\,.
\label{spatlim15}
\end{equation}
The transformation \eqref{spatlim15} will also later allow us to
write the averaged equations in the synchronous gauge for the
averaged geometry.

We now proceed to calculating the correlation $2$-form
$\bZ{a}{b}{i}{j}$ and thereby the averaged Einstein equations.
\label{spatlim}

\section{The averaged cosmological field equations}
\noindent We start by defining (in any gauge with $N^A=0$) the
expansion tensor $\Theta{}^A_B$ by
\begin{equation}
\Theta{}^A_B\equiv \frac{1}{2N}h^{AC}\dot h_{CB}\,,
\label{correln1}
\end{equation}
where the dot will always refer to a derivative with respect to
the VPC time $\bt$, and $h^{AB}$ is the inverse of the $3$-metric
$h_{AB}$. (This also gives the symmetric tensor $\Theta_{AB} =
(1/2N)\dot h_{AB}$, which is the negative of the extrinsic
curvature tensor.) The traceless symmetric shear tensor $\sigma{}^A_B$
and the shear scalar $\sigma^2$ are defined by
\begin{equation}
\sigma{}^A_B\equiv \Theta{}^A_B-(\Theta/3) \delta{}^A_B ~~;~~
\sigma^2 \equiv \frac{1}{2}\sigma{}^A_B\sigma{}^B_A\,,
\label{correln2}
\end{equation}
where $\Theta\equiv\Theta{}^A_A = (1/N)\partial_{\bt}\ln{\sqrt h}$ is
the expansion scalar. 

The connection $1$-forms $\om^i_{\ph{i}j} = \Gamma{}^i_{jk}\dx{k}$
can be easily calculated in terms of the expansion tensor, for an
arbitrary lapse function $N$. Specializing to the volume
preserving gauge ($N=h^{-1/2}$), the bilocal extensions
$\Om{i}{j}$ of the connection $1$-forms are trivial and are simply
given by
\begin{equation}
\Om{i}{j}(x^\prime,x) = \Gamma{}^i_{jk}(x^\prime)\dx{k}\,.
\label{correln6}
\end{equation}
Since $G_{ab} = \bar g_{ab}$, the connection $1$-forms
$\barOm{i}{j}$ for the averaged manifold \Mbar\ are constructed
using the FLRW metric in volume preserving gauge given in
\eqn{spatlim11}, and can also be easily evaluated.

We can now construct the correlation $2$-form $\bZ{a}{b}{i}{j}$
and from there the correlation tensor :
\begin{equation}
C{}^a_b = \left(Z^a_{\ph{a}ijb} - \frac{1}{2}\delta{}^a_b
Z^m_{\ph{a}ijm} \right)G^{ij}. \label{correln100}
\end{equation}
Now, the components of the Einstein tensor $E{}^a_b$ for the
averaged spacetime with metric \eqref{spatlim11} are given by
\begin{align}
E{}^{\bt}_{\bt} &= 3\ba^6H^2 ~~;~~ E{}^{\bt}_A = 0 = E{}^B_{\bt}
\,,
\nonumber\\
E{}^A_B &= \ba^6\delta{}^A_B\left[ 2\left(\frac{\ddot\ba}{\ba} +
  3H^2\right) + H^2\right] \,,
\label{correln11}
\end{align}
where the peculiar splitting of terms in the last equation is for
later convenience. Recall that the overdot denotes a derivative
with respect to the VPC time $\bt$, not synchronous time. In terms
of the coordinate independent objects introduced in
\eqn{spatlim-FLRW1}, we have
\begin{equation}
j_1(x) = -3\ba^6H^2 ~~;~~ j_2(x) = \ba^6\left[
  2\left(\frac{\ddot\ba}{\ba} + 3H^2\right) + H^2\right] \,.
\label{correln-FLRW1}
\end{equation}
From the averaged Einstein equations  we next
construct the scalar equations which in the standard case would
correspond to the Friedmann equation and the Raychaudhuri
equation. These correspond to the Einstein tensor components,
\begin{equation}
E{}^a_b\bar v^b\bar v_a  = j_1(x)  ~~;~~ \pi^b_aE{}^a_b +
E{}^a_b\bar v^b\bar v_a = 3j_2(x) + j_1(x)\,, \label{correln12}
\end{equation}
and are given by
\begin{subequations}
\begin{align}
&3\ba^6H^2 =\left(\kappa T{}^a_b - C{}^a_b\right)\bar v_a\bar v^b
  \nonumber\\ 
&~~~~~~~~~=\kappa\bar\rho - \frac{1}{2}\left[ \Cal{Q}^{(1)} +
  \Cal{S}^{(1)} \right]  \,,
\label{correln13a} \\&\nonumber\\
&6\ba^6\left( \frac{\ddot\ba}{\ba} + 3H^2 \right) = \left(-\kappa 
T{}^a_b + C{}^a_b\right) \left(\bar v_a\bar v^b + \pi^b_a \right)
\nonumber\\ 
&~~~~~~~~~~= -\kappa\left(\bar\rho+3\bar p\right) + 2\left[
  \Cal{Q}^{(1)} +  \Cal{Q}^{(2)} + \Cal{S}^{(2)} \right]\,.
\label{correln13b}
\end{align}
\label{correln13}
\end{subequations}
\noindent Here \eqn{correln13a} is the modified Friedmann equation
and \eqn{correln13b} the modified Raychaudhuri equation (in the
vol0ume preserving gauge on \Mbar). We have used
\eqn{spatlim-T-ab1}, with the overbar on $\rho$ and $p$ reminding
us that they are expressed in terms of the nonsynchronous time
$\bt$, and we have defined the correlation terms
\begin{subequations}
\begin{align}
\Cal{Q}^{(1)} &=
\ba^6\left[\frac{2}{3}\left(\avg{\frac{1}{h}\Theta^2} -
\frac{1}{\ba^6}(^{\rm F}\Theta^2)\right) -
2\avg{\frac{1}{h}\sigma^2}\right] ~; \nonumber\\
&~~~~~~~~~~~~~~~~~~~~~~~~~~~~~~~~ \frac{1}{\ba^6}(^{\rm F}\Theta^2)
= \left(3H\right)^2\,, 
\label{correln14a} \\&\nonumber\\
\Cal{S}^{(1)} &= \frac{1}{\ba^2}\delta^{AB}\left[
  \avg{\,^{(3)}\Gamma{}^J_{AC}\,^{(3)}\Gamma{}^C_{BJ}}
  \right. \nonumber\\ 
&~~~~~~~~~~~~~~~~~~\left. -
  \avg{\partial_A(\ln\sqrt{h})\partial_B(\ln\sqrt{h})}
   \right] \,,
\label{correln14b} \\&\nonumber\\
\Cal{Q}^{(2)} &= \ba^6\avg{\frac{1}{h}\Theta{}^A_B\Theta{}^B_A} -
\frac{1}{\ba^2}\delta^{AB}\avg{\Theta_{AJ}\Theta{}^J_B},
\label{correln14c} \\&\nonumber\\
\Cal{S}^{(2)} &= \ba^6\avg{\frac{1}{h}h^{AB}
  \partial_A(\ln\sqrt{h})\partial_B(\ln\sqrt{h})} \nonumber\\ 
&~~~~~~~~~~~-  \frac{1}{\ba^2}\delta^{AB}
  \avg{\partial_A(\ln\sqrt{h})\partial_B(\ln\sqrt{h})}  \,.
\label{correln14d}
\end{align}
\label{correln14}
\end{subequations}
%
In defining $\Cal{Q}^{(1)}$ we have used the relation $\Theta^2 -
\Theta{}^A_B\Theta{}^B_A = (2/3)\Theta^2 - 2\sigma^2$.
$\Cal{Q}^{(1)}$ and $\Cal{Q}^{(2)}$ are correlations of the
extrinsic curvature, whereas $\Cal{S}^{(1)}$ and $\Cal{S}^{(2)}$
are correlations restricted to the intrinsic $3$-geometry of the
spatial slices of \Cal{M}. Since the components of $C{}^a_b$ are
not explicitly constrained we can treat the combinations
$(1/2)(\Cal{Q}^{(1)} + \Cal{S}^{(1)})=-C{}^0_0$ and
$2(\Cal{Q}^{(1)} + \Cal{Q}^{(2)} + \Cal{S}^{(2)}) =
(C{}^A_A-C{}^0_0)$ as independent, subject only to the
differential constraints which we will come to
below.

As discussed in the beginning of Section \ref{spatlim}, the
remaining components of $C{}^a_b$ must be set to zero, giving
constraints on the underlying inhomogeneous geometry. In
coordinate independent language, these constraints read
\begin{align}
\pi^b_k C{}^a_b\bar v_a &= 0 =\,\pi^k_a C{}^a_b\bar v^b ~;\nonumber\\ 
\pi^i_a\pi^b_k C{}^a_b &- \frac{1}{3}\pi^i_k \left(\pi^b_a
C{}^a_b\right) = 0\,. \label{correln15}
\end{align}
\eqns{correln15} reduce to the following for our specific choice
of volume preserving coordinates,
\begin{equation}
C{}^0_A = 0 ~~;~~ C{}^A_0 = 0 ~~;~~ C{}^A_B -
\frac{1}{3}\delta{}^A_B(C{}^J_J) = 0 \,, \label{correln16}
\end{equation}
It can be shown that the VPC assumption $N=h^{-1/2}$ reduces the
correlations $\Cal{Q}^{(2)}$ and $\Cal{S}^{(2)}$ defined in
\eqns{correln14c} and \eqref{correln14d}, as well as several terms
in the explicit expansion of \eqn{correln16}, to the form
\begin{equation}
\frac{1}{\avg{g_{00}}}\avg{g_{00}g^{AB}\Gamma{}^{a_1}_{b_1c_1}
  \Gamma{}^{i_1}_{j_1k_1}} - \avg{g^{AB}}\avg{\Gamma{}^{a_2}_{b_2c_2}
  \Gamma{}^{i_2}_{j_2k_2}}\,.
\label{correln18}
\end{equation}
It can be shown that
\begin{align}
\avg{g_{00}g^{AB}\Gamma{}^a_{bc}\Gamma{}^i_{jk}} &=
\avg{g_{00}g^{AB}}\avg{\Gamma{}^a_{bc}\Gamma{}^i_{jk}} \nonumber\\ 
&= -\avg{\frac{h^{AB}}{h}}\avg{\Gamma{}^a_{bc}\Gamma{}^i_{jk}} \,.
\label{correln19}
\end{align}
An interesting point is that the VPC assumption $N=h^{-1/2}$
further allows us to assume $\avg{h^{AB}/h} =
\avg{h^{AB}}\avg{1/h}$ consistently with the formalism. Using \eqn{spatlim10} this gives us
\begin{equation}
\avg{\frac{h^{AB}}{h}} = \frac{1}{\ba^6}\avg{h^{AB}}\,.
\label{correln20}
\end{equation}
This  shows that the correlation terms $\Cal{Q}^{(2)}$ and
$\Cal{S}^{(2)}$ in fact vanish,
\begin{equation}
\Cal{Q}^{(2)} = 0 = \Cal{S}^{(2)}\,, \label{correln21}
\end{equation}
and leads to some remarkable cancellations in \eqns{correln16},
which simplify to give
\begin{subequations}
\begin{align}
&\delta^{JK}\left[\avg{\sqrt{h}\Theta_{JB}\,^{(3)}\Gamma{}^B_{AK}}
--
  \avg{\sqrt{h}\Theta_{JK}\,^{(3)}\Gamma{}^B_{AB}} \right] = 0\,,
\label{correln22a}\\
&\delta^{JK}\avg{\frac{1}{\sqrt
h}\Theta{}^B_K\,^{(3)}\Gamma{}^A_{JB}}
  - \delta^{AJ}\avg{\frac{1}{\sqrt
      h}\Theta{}^K_K\,^{(3)}\Gamma{}^B_{JB}} = 0\,,
\label{correln22b}\\
&\delta^{JK}\avg{\,^{(3)}\Gamma{}^A_{JC}\,^{(3)}\Gamma{}^C_{KB}} -
\delta^{AJ}\avg{\,^{(3)}\Gamma{}^C_{JC}\,^{(3)}\Gamma{}^K_{BK}}
\nonumber\\ 
&~~~~~~~~~~~~~~~~~~~~~~~~~~~~~~~~~~~~~~~~~~~~~=
\frac{1}{3}\delta{}^A_B\left(\ba^2\Cal{S}^{(1)}\right) \,.
\label{correln22c}
\end{align}
\label{correln22}
\end{subequations}
These simplifications are solely a consequence of assuming that
the inhomogeneous metric in the volume preserving gauge averages
out to give the FLRW metric in standard form. In general, these
simplifications will not occur when the standard FLRW metric
arises from an arbitrary choice of gauge for the inhomogeneous
metric.

In order to come as close as possible to the standard approach in
Cosmology, we will now rewrite the scalar equations
\eqref{correln13} (which are the cosmologically relevant ones)
after performing the transformation given in \eqn{spatlim15} in
order to get the FLRW metric to the form
\begin{equation}
^{(\Mbar)}ds^2 = -d\tau^2 + a^2(\tau)\delta_{AB}dy^Ady^B ~~;~~
a(\tau)
  =  \ba(\bt(\tau))\,.
\label{correln23}
\end{equation}
Since \eqns{correln13} are \emph{scalar} equations, this
transformation only has the effect of reexpressing all the terms
as functions of the synchronous time $\tau$. Although the
transformation will change the explicit form of the coordination
bivector $\W{a}{j}$, this change involves only the time
coordinate, and in the spatial averaging limit there is no
difference between averages computed in the VPCs and those
computed after the time redefinition. This again emphasizes the
importance of the spatial averaging limit of spacetime averaging,
if we are to succeed operationally in explicitly displaying the
correlations as corrections to the standard cosmological
equations. The correlation terms in \eqns{correln14} are therefore
still interpreted with respect to the volume preserving gauge, but
are treated as functions of $\tau$. For the scale factor on the
other hand, we have
\begin{equation}
\ba^3H = \frac{1}{a}\frac{da}{d\tau} \equiv H_{\rm FLRW} ~~;~~
\ba^6\left(\frac{\ddot\ba}{\ba}+3H^2\right) =
\frac{1}{a}\frac{d^2a}{d\tau^2} \,. \label{correln24}
\end{equation}
Further writing
\begin{equation}
\rho(\tau) = \bar\rho(\bt(\tau))  ~~;~~ p(\tau) = \bar
p(\bt(\tau)) \,, \label{correln25}
\end{equation}
equations \eqref{correln13} become
\begin{subequations}
\begin{align}
H^2_{\rm FLRW} &= \frac{8\pi G_N}{3}\rho -  \frac{1}{6}\left[
  \Cal{Q}^{(1)} + \Cal{S}^{(1)} \right],
\label{correln26a} \\&\nonumber\\
\frac{1}{a}\frac{d^2a}{d\tau^2} &= -\frac{4\pi G_N}{3}\left(\rho +
  3p\right) + \frac{1}{3}\Cal{Q}^{(1)} \,.
\label{correln26b}
\end{align}
\label{correln26}
\end{subequations}
We emphasize that the quantities $\Cal{Q}^{(1)}$ and
$\Cal{S}^{(1)}$, defined in \eqns{correln14a} and
\eqref{correln14b} as correlations in the \emph{volume preserving}
gauge, are to be thought of as functions of the \emph{synchronous}
time $\tau$, where the coordinate $\tau$ itself was defined
\emph{after} the spatial averaging. Such an identification is
justified since we are dealing with scalar combinations of these
quantities. Note that $\Cal{Q}^{(1)}$ and $\Cal{S}^{(1)}$ can be
treated independently, apart from the constraints imposed by
conservation conditions, which we turn to next. These conservation
conditions can be decomposed into a scalar part and a $3$-vector
part, given respectively by
\begin{equation}
\ba-r v^bC{}^a_{b;a} = 0 ~~;~~ \pi^b_k C{}^a_{b;a} = 0 \,.
\label{correln27}
\end{equation}
In the synchronous gauge \eqref{correln23} for the FLRW metric,
the scalar equation reads
\begin{equation}
\left(\partial_\tau\Cal{Q}^{(1)} + 6H_{\rm
FLRW}\Cal{Q}^{(1)}\right) + \left(\partial_\tau\Cal{S}^{(1)} +
2H_{\rm FLRW}\Cal{S}^{(1)}\right) = 0 \,. \label{correln28}
\end{equation}
We recall that this equation is a consequence of setting the
correlation 3-form and the correlation 4-form to zero, and it
relates the evolution of $\Cal{Q}^{(1)}$ and $\Cal{S}^{(1)}$. The
$3$-vector equation (on imposing the first set of conditions in
\eqn{correln15}) simply gives $\partial_\tau C{}^\tau_A = 0$, so
that $C{}^\tau_A = 0 =\,$constant, which also implies that
$C{}^A_\tau = 0=\,$constant and hence this equation gives nothing
new. (We have used the relations $C{}^0_0 = C{}^\tau_\tau$,
$C{}^0_A = \ba^3 C{}^\tau_A$ and $C{}^A_0 = (1/\ba^3)C{}^A_\tau$
where $0$ denotes the nonsynchronous time coordinate $\bt$.)

The cosmological equations (\ref{correln26}), along with the
constraint equations (\ref{correln22}) and (\ref{correln28}) are
the key results of this section. Subject to the acceptance of the
volume preserving gauge on the underlying manifold ${\cal M}$ they
can in principle be used to study the role of the correction terms
resulting from spatial averaging.

\subsection{A comparison with the  averaging formalism of Buchert}
\noindent The averaging formalism developed by Buchert is based
exclusively on the manifold ${\cal M}$, and there is no analog of
the averaged manifold ${\cal \bar{M}}$ in this scheme. Given an
inhomogeneous metric on ${\cal M}$ one takes the trace of the
Einstein equations in the {\it inhomogeneous} geometry, and
carries out a spatial averaging of the inhomogeneous scalar
equations.

For ease of comparison, we again recall in brief Buchert's construction, by
first writing down the averaged equations for the simplest case of
pressureless and irrotational inhomogeneous dust. The metric can
be written in synchronous and comoving gauge as
\begin{equation}
ds^2=\,-dt^2+b_{AB}(\rmb{x},t)dx^Adx^B\,. \label{avg110}
\end{equation}
The Einstein equations can be split  into a set of
scalar equations and a set of vector and traceless tensor
equations. The scalar equations are the Hamiltonian constraint
\eqref{avg22a} and the evolution equation for $\Theta$
\eqref{avg22b},
\begin{subequations}
\begin{equation}
\Cal{R}+\frac{2}{3}\Theta^2-2\sigma^2=16\pi G\rho\,, \label{avg22a}
\end{equation}
\begin{equation}
\Cal{R}+\partial_t\Theta+\Theta^2=12\pi G\rho\,, \label{avg22b}
\end{equation}
\label{avg22}
\end{subequations}
where \Cal{R}\ is the Ricci scalar of the 3-dimensional
hypersurface of constant $t$, $\Theta$ and $\sigma^2$ are the
expansion scalar and the shear scalar defined earlier and $\rho$
is the inhomogeneous matter density of the dust. Note that all
quantities in Eqns. \eqref{avg22} generically depend on both
position $\rmb{x}$ and time $t$. Eqns. \eqref{avg22a} and
\eqref{avg22b} can be combined to give Raychaudhuri's equation
\begin{equation}
\partial_t\Theta+\frac{1}{3}\Theta^2+2\sigma^2+4\pi G\rho=0\,.
\label{avg33}
\end{equation}
The continuity equation $\partial_t\rho=-\Theta\rho$ which gives
the evolution of $\rho$, is consistent with Eqns. \eqref{avg22a},
\eqref{avg22b}. Only scalar Einstein equations are considered,
since the spatial average of a scalar quantity can be defined in a
gauge covariant manner, within a given foliation of space-time. We
return to this point below. For the space-time described by
\eqref{avg110}, the spatial average of a scalar $\Psi(\rmb{x},t)$
over a \emph{comoving} domain \Cal{D} at time $t$ is defined by
\begin{equation}
\avgD{\Psi}=\frac{1}{\uD{V}}\int_\Cal{D}{d^3x\sqrt{b}\,\Psi}\,,
\label{avg44}
\end{equation}
where $b$ is the determinant of the 3-metric $b_{AB}$ and $\uD{V}$
is the volume of the comoving domain given by
$\uD{V}=\int_\Cal{D}{d^3x\sqrt{b}}$. Spatial averaging is, by
definition, not generally covariant. Thus the choice of foliation
is relevant, and should be motivated on physical grounds. In the
context of cosmology, averaging over freely-falling observers is a
natural choice, especially when one intends to compare the results
with standard FLRW cosmology.  Following the definition
\eqref{avg44} the following commutation relation then holds
\begin{equation}
\partial_t\avgD{\Psi}-\avgD{\partial_t\Psi}=
\avgD{\Psi\Theta}-\avgD{\Psi}\avgD{\Theta}\,, \label{avg55}
\end{equation}
which yields for the expansion scalar $\Theta$
\begin{equation}
\partial_t\avgD{\Theta}-\avgD{\partial_t\Theta}=
\avgD{\Theta^2}-\avgD{\Theta}^2\,. \label{avg66}
\end{equation}
Introducing the dimensionless scale factor
$\aD\equiv\left(\uD{V}/V_{\Cal{D} i}\right)^{1/3}$ normalized by
the volume of the domain \Cal{D}\ at some initial time $t_i$, we
can average the scalar Einstein equations \eqref{avg22a},
\eqref{avg22b} and the continuity  equation to obtain
\begin{subequations}
\begin{equation}
\partial_t\avgD{\rho}=\,-\avgD{\Theta}\avgD{\rho} ~~~;~~~
\avgD{\Theta}=3\frac{\partial_t\aD}{\aD}\,, \label{avg77a}
\end{equation}
\begin{equation}
\left(\frac{\partial_t\aD}{\aD}\right)^2=\frac{8\pi
G}{3}\avgD{\rho} -
\frac{1}{6}\left(\uD{\Cal{Q}}+\avgD{\Cal{R}}\right)\,,
\label{avg77b}
\end{equation}
\begin{equation}
\left(\frac{\partial_t^2\aD}{\aD}\right)= -\frac{4\pi G}{3}
\avgD{\rho}+ \frac{1}{3}\uD{\Cal{Q}}\,. \label{avg77c}
\end{equation}
\label{avg77}
\end{subequations}
Here, the `kinematical backreaction' $\uD{\Cal{Q}}$ is given by
\begin{equation}
\uD{\Cal{Q}}\equiv\frac{2}{3}\left(\avgD{\Theta^2}-
\avgD{\Theta}^2\right)-2\avgD{\sigma^2} \label{avg88}
\end{equation}
and is a spatial constant over the domain \Cal{D}.

 A necessary condition for \eqref{avg77c} to integrate to
\eqref{avg77b} takes the form of the following differential
equation involving $\uD{\Cal{Q}}$ and $\avgD{\Cal{R}}$,
\begin{equation}
\partial_t\uD{\Cal{Q}}+6\frac{\partial_t\aD}{\aD}\uD{\Cal{Q}}
+\partial_t\avgD{\Cal{R}}+2\frac{\partial_t\aD}{\aD}\avgD{\Cal{R}}=
0\,. \label{avg99}
\end{equation}

The equations above describe the essence of Buchert's averaging
formalism, for the dust case. We note that the remaining eight
Einstein equations for the inhomogeneous geometry, which are not
scalar equations, are not averaged. These are the five evolution
equations for the trace-free part of the shear,
\begin{equation}
\partial_t\left(\sigma{}^A_B\right) = - \Theta
\sigma{}^A_B -  \Cal{R}{}^A_B +\frac{2}{3}\delta{}^A_B
\left(\sigma^2 - \frac{1}{3}\Theta^2 + 8\pi G\rho \right) \,.
\label{shea1}
\end{equation}
and the three equations relating the spatial variation of the
shear and the expansion,
\begin{equation}
\sigma{}^A_{B || A} = \frac{2}{3} \Theta_{|| B}\,. \label{shea2}
\end{equation}
Here, $\Cal{R}{}^A_B$ is the spatial Ricci tensor and, in
Buchert's notation, a $||$ denotes covariant derivative with
respect to the $3$-metric.

In analogy with the dust case, Buchert's averaging formalism can
be applied to the case of a perfect fluid, by
starting from the metric
\begin{equation}
ds^2 = -N^2 dt^2 + b_{AB}dx^A dx^B \,. \label{Nmet}
\end{equation}
The averaged scalar Einstein equations for the scale factor \aD\
are
\begin{equation}
3\frac{\partial_t^2 \aD}{\aD} + 4\pi G
\avgD{N^2\left(\rho+3p\right)}= \uD{\bar{\Cal{Q}}} +
\uD{\bar{\Cal{P}}} \,, \label{ep1}
\end{equation}
\begin{equation}
6 \uD{H}^2 - 16\pi G \avgD{N^2\rho}= -\uD{\bar{\Cal{Q}}} -
\avgD{N^2\Cal{R}}~~;~~ \uD{H}=\frac{\partial_t\aD}{\aD}\,,
\label{ep2}
\end{equation}
where the kinematical backreaction $\uD{\bar{\Cal{Q}}}$ is given
by
\begin{equation}
\uD{\bar{\Cal{Q}}} = \frac{2}{3}\left(
\avgD{\left(N\Theta\right)^2} - \avgD{N\Theta}^2\right)
 - 2\avgD{N^2\sigma^2}\,,
\label{kb}
\end{equation}
and the dynamical backreaction $\uD{\bar{\Cal{P}}}$ is given by
\begin{equation}
\uD{\bar{\Cal{P}}} = \avgD{N^2\Cal{A}} + \avgD{\Theta\partial_tN}
\,, \label{db}
\end{equation}
where $\Cal{A}=\nabla_j(u^i\nabla_iu^j)$ is the $4$-divergence of
the $4$-acceleration of the fluid. \eqn{ep2} follows as an
0integral from \eqn{ep1} if and only if the relation
\begin{align}
\partial_t \uD{\Cal{Q}} &+ 6\uD{H}\uD{\Cal{Q}} +
\partial_t \avgD{N^2\Cal{R}} + 2\uD{H}\avgD{N^2\Cal{R}} +
4\uD{H}\uD{\bar{\Cal{P}}} \nonumber\\
&- 16\pi G \left[\partial_t \avgD{N^2\rho} +
3\uD{H}\avgD{N^2\left(\rho+p\right)} \right]  =0\,, \label{incon}
\end{align}
is satisfied. There are also the unaveraged equations (which we do
not display here) for the shear, analogous to the shear equations
\eqref{shea1} and \eqref{shea2} for dust.

Buchert's approach is the only other approach, apart from
Zalaletdinov's MG, which is capable of treating inhomogeneities in
a nonperturbative manner, although it is limited to using only
scalar quantities within a chosen $3+1$ splitting of spacetime.
Buchert takes the trace of the Einstein equations in the
\emph{inhomogeneous} geometry, and averages these inhomogeneous
scalar equations. In the context of Zalaletdinov's MG however, we
have used the existence of the vector field $\bar v^a$ in the FLRW
spacetime to construct scalar equations \emph{after} averaging the
full Einstein equations. As far as observations are concerned, it
has been noted by Buchert and Carfora that the
spatially averaged matter density $\avgD{\rho}$ defined by Buchert
is \emph{not} the appropriate observationally relevant quantity --
the ``observed'' matter density (and pressure) is actually defined
in a \emph{homogeneous} space. Since we have done precisely this
in \eqn{spatlim-T-ab1}, we are directly dealing with the
appropriate observationally relevant quantity in the MG framework.

Another important difference between the two approaches is the
averaging operation itself. Buchert's spatial average, defined for
scalar quantities, is given (for some scalar $\Psi(t,x^A)$) by
(\ref{avg44}) above. On the other hand the averaging operation we
have been using (given by \eqn{spatlim8} using the volume
preserving gauge) is a limit of a spacetime averaging defined
using the coordination bivector $\W{a}{j}$, and is different from
the one in \eqn{avg44}.

Most importantly though, Buchert's averaging scheme by itself does
not incorporate the concept of an averaged manifold \Mbar\
(although the work of Buchert and Carfora \cite{Buchert} does
deal with $3$-spaces of constant curvature). In a recent paper we had argued that Buchert's ``effective scale
factor''
$a_\Cal{D}(t)\equiv(V_\Cal{D}(t)/V_\Cal{D}(t_{in}))^{1/3}$ must be
the scale factor for the metric of the averaged manifold, upto
some corrections arising due to such effects as calculated by
Buchert and Carfora. In the present work however, it is clear that
such a suggestion is necessarily incomplete due to the presence of
\eqns{correln22} constraining the underlying geometry. These
constraints are in general nontrivial and hence indicate that it
is not sufficient to assume that the metric of the inhomogeneous
manifold averages out to the FLRW form -- there are additional
conditions which the correlations must satisfy.

To our understanding, Buchert's averaging formalism is a valid
aproach, even though it is based on a spatial averaging. A
central difference from the MG approach is the issue of closure :
not all the Einstein equations have been averaged in Buchert's
approach, but only the scalar ones. This puts a constraint on the
allowed solutions considered for the averaged equations:
\eqref{avg77} for the dust case, and \eqref{ep1} and \eqref{ep2}
for the fluid case. Solutions to these equations must necessarily
be checked for consistency with the unaveraged equations for the
shear. Further, averaging over successively larger scales can
bring in additional corrections to the averaged equations, as
discussed by Buchert and Carfora. Also, if one does not wish to
identify Buchert's \aD\ with the scale factor in FLRW cosmology,
one is compelled to develop a whole new set of ideas in order to
try and compare theory with observation. On the other hand, if one
does identify \aD\ with the scale factor, comparison with standard
cosmology becomes more convenient, but this brings in additional
constraints on the underlying inhomogeneous geometry. Thus our
conclusion is that the Buchert formalism is a correct and
tractable averaging scheme, provided all the caveats pointed out
in this paragraph are taken care of. Also, when these caveats have
been taken care of correctly, the Buchert formalism is expected to
give the same physical results as the MG approach. We recall that
in the covariant MG approach also, once a spacetime geometry has
been identified for the averaged manifold \Mbar, a gauge must be
selected for the geometry on the underlying manifold, in order to
explicitly compute the correction scalars for comparison with
observation.

The advantage of the MG approach is that it accomplishes in a neat
package what the Buchert approach, with its attendant caveats,
sets out to do. In the MG approach, there are no unaveraged shear
equations, because the trace of the Einstein equations has been
taken after performing the averaging on the underlying geometry.
Since the averaged geometry is FLRW, the shear is zero by
definition. There is a natural metric on the averaged manifold by
construction, the FLRW metric. The correlations satisfy additional
constraints, given by Eqns. (\ref{correln22}). Thus, once a gauge
has been chosen and if one can overcome the computational
complexity of the averaging operation, the cosmological equations
derived by us in the MG approach are complete and ready for
application, without any further caveats.

In spite of these differences, our equations \eqref{correln26} and
\eqref{correln28} for the volume preserving gauge are strikingly
similar to Buchert's effective FLRW equations and their
integrability condition in the dust case; and in the case of
general $N$, the role of Buchert's dynamical backreaction
$\uD{\bar{\Cal{P}}}$ in \eqns{ep1} and \eqref{incon} is identical
to that of our combination of
$(\tilde{\Cal{P}}^{(2)}+\tilde{\Cal{S}}^{(2)})$. Concentrating on the volume preserving case,
the structure of the correlation $\Cal{Q}^{(1)}$ is identical to
Buchert's kinematical backreaction $\uD{\Cal{Q}}$ (or
$\uD{\bar{\Cal{Q}}}$ in the general case). The correlation
$\Cal{S}^{(1)}$ appears in place of the averaged $3$-Ricci scalar
$\avgD{\Cal{R}}$ in Buchert's dust equations. This is not
unreasonable since Buchert's $\avgD{\Cal{R}}$ can be thought of as
$\avgD{\Cal{R}} = 6k_\Cal{D}/a_\Cal{D}^2 +\,$corrections, where
$6k_\Cal{D}/a_\Cal{D}^2$ represents the $3$-Ricci scalar on the
averaged manifold which in our case is zero, and hence
$\Cal{S}^{(1)}$ represents the corrections due to averaging.
Further, these similarities are in spite of the fact that our
correlations were defined assuming that a \emph{volume preserving}
gauge averages out to the FLRW $3$-metric in standard form,
whereas Buchert's averaging is most naturally adapted to beginning
with a \emph{synchronous} gauge. This remarkable feature, at least
to our understanding, does not seem to have any deeper meaning --
it simply seems to arise from the structure of the Einstein
equations themselves, together with our assumption
$\rmb{D}_{\bar\Omega}\bZ{a}{b}{i}{j} = 0$. In the absence of this
latter condition, one would have to consider the correlation $3$-
and $4$-forms mentioned earlier, and the structure of the
correlation terms and their ``conservation'' equations would be
far more complicated.

An entirely different outlook towards his approach has been 
emphasized to us by Buchert. According to Buchert, the  
absence of an averaged manifold \Mbar\ is not to be thought of as a
`caveat', but as a feature deliberately retained `on purpose'. The
actual inhomogeneous Universe is regarded by Buchert as the only
fundamental entity, and the introduction of an averaged Universe is in
fact regarded as an unphysical and unnecessary approximation. As we
mentioned earlier, this is probably the most important difference
between MG and Buchert's approach. In the latter, contact with
observations is to be made by constructing averaged quantities, such
as the scalars defined earlier in this section, and by introducing the
expansion factor $\aD$. The assertion here is that the averaging of
\emph{geometry}, as discussed in MG or in the Renormalization Group
approach of Buchert and Carfora \cite{Buchert} is not an
indispensable step in comparing the inhomogeneous Universe with actual
observations. The need for averaging of geometry is to be physically
separated from simply looking at effective properties (such as the
constructed scalars) which can be defined for any inhomogeneous
metric. Averaging of geometry becomes relevant if (i) an observer
insists on interpreting the data in a FLRW template model, so that
(s)he needs a mapping from the actual inhomogeneous slice and its
average properties to the corresponding properties in this template,
or (ii) one desires a mock metric, to sort of have a thermodynamic
effective metric to approximate the real one.  In this context it
should perhaps also be mentioned that the importance of a thin
time-slice approximation of spacetime averaging (as opposed to a
strict spatial averaging) has been stressed also by Buchert.

\section {Perturbation theory, structure formation, and backreaction}

We have in hand the machinery to ask the following question : Is cosmological perturbation theory stable against growth of backreaction? The answer must be found iteratively. Assume a background with perturbations on it, calculate the back-reaction, feed it in the right hand of the modified Friedmann equations to find the new background, and so on :
\be
a^{(0)} \rightarrow \phi^{(0)} \rightarrow C^{(0)}
\rightarrow a^{(1)} \rightarrow \phi^{(1)} \rightarrow
\ldots 
\label{MG10}
\ee
Let the perturbed FLRW metric be 
\be ds^{2}=
a^2\left[ -(1+2\phi)d\eta^2 + 2\om_Adx^Ad\eta +
\left((1-2\psi)\gamma_{AB} +  \chi_{AB}\right)dx^Adx^B \right]\,.
\label{linPT1}
\ee
We work with a VPC which has no residual degrees of freedom. Further, this VPC is constructed by starting from the conformal Newtonian gauge, and by making a steady coordinate transformation. This ensures that all averaged quantities are gauge invariant.
We evaluate the correlation scalars for a given initial power spectrum - standard CDM.

For a constant nonevolving potential $\phi(\vec{x})$, and with a power spectrum
\be
\frac{k^3P_{\phi i}(k)}{2\pi^2} = A (k/H_0)^{n_s - 1}\,,
\label{exs-eds3}
\ee
the back reaction is
\be
\frac{\Cal{S}^{(1)}}{H_0^2} \sim -\frac{1}{a^2}(10^{-4})\,.
\label{exs-eds5}
\ee
The smallness of backreaction holds also for the exact sCDM model thus demonstrating the stability of perturbation theory against the growth of back-reaction.

This analysis ignores contribution of scales that have become fully nonlinear in matter density at late times and it is important to ask if structure formation can significantly modify large sacle dynamics. 

We studied backreaction in a toy model of spherical collapse, using the LTB solution.
The initial density is chosen to be
\begin{equation}
\rho(t_i,r)=\rho_{bi}\left\{
\begin{array}{l}
(1+\delta_\ast),~~~~r<r_\ast\\
(1-\delta_v),~~~r_\ast<r<r_v\\
1,~~~~~~~~~~~~r>r_v\,,
\end{array}\right .
\label{nonlin-2eq6}
\end{equation}

We match the initial velocity and coordinate scaling to the global
background solution, by requiring 
\begin{align}
R(t_i,r) &= a_i r\,,
\label{nonlin-2eq8} \\
\dot R(t_i,r) &= a_i H_i r \,,
\label{nonlin-2eq9}
\end{align}

For the FLRW background we consider an Einstein-deSitter (EdS)
solution with scale factor and Hubble parameter given by 
\begin{align}
&a(t) = (t/t_0)^{2/3} ~~;~~ t_0 = 2/(3H_0) \,,
\label{nonlin-2eq10} \\
&H(t) \equiv \dot a/a = 2/(3t)\,,
\label{nonlin-2eq11}
\end{align}
with $t_0$ denoting the present epoch. $a_i$ fixes the initial time as 
\begin{equation}
t_i = 2/(3H_0) a_i^{3/2}\,.
\label{nonlin-2eq12}
\end{equation}
We use $a_i=10^{-3}$, so that the initial conditions are
being set around the CMB last scattering epoch.
The mass function $M(r)$ and curvature function $k(r)$  in this LTB solution are given by
\begin{equation}
GM(r)=\frac{1}{2}H_0^2r^3\left\{ 
\begin{array}{l}
1 + \delta_\ast,~~~~0<r<r_\ast\\
1 + \delta_v\left( \left(r_c/r\right)^3 - 1 \right),
~~~r_\ast<r<r_v\\    
1+(\delta_v/r^3)\left(r_c^3 - r_v^3\right),~~r>r_v\,, 
\end{array}\right .
\label{nonlin-2eq14}
\end{equation}
where we have defined a ``critical'' radius $r_c$ by the equation
\begin{equation}
\left(\frac{r_c}{r_\ast}\right)^3 = 1 + \frac{\delta_\ast}{\delta_v} 
\,. 
\label{nonlin-2eq15}
\end{equation}
The significance of $r_c$  is brought out by $k(r)$ :
\begin{equation}
k(r)=\frac{H_0^2}{a_i}\left\{ 
\begin{array}{l}
\delta_\ast,~~~~r<r_\ast\\
\delta_v\left(\left(r_c/r\right)^3-1\right), ~~~ r_\ast<r<r_v\\    
(\delta_v/r^3)\left(r_c^3 - r_v^3\right), ~~ r>r_v\,.  
\end{array}\right .
\label{nonlin-2eq17}
\end{equation}
Since
$\delta_\ast,\delta_v>0$, we have $r_c>r_\ast$ by definition. The following possibilities arise :

If $r_c>r_v$, then $k(r)>0$ for all $r$, and every shell will 
ultimately collapse, including the ``void'' region $r_\ast<r<r_v$. 
If $r_c<r_v$, then $k(r)>0$ for $r<r_c$ and changes sign at
  $r=r_c$. Hence, the region $r_\ast<r<r_c$ will collapse even
  though it is underdense, while the region $r>r_c$ will expand
  forever.  
If $r_c=r_v$, then the ``void'' exactly compensates for the
  overdensity, and the universe is exactly EdS for $r>r_v$.

Transforming to the perturbed FLRW form :
We want a coordinate
transformation $(t,r)\to(\tau,\tilde{r})$ such that the metric in the
new coordinates is
\begin{equation}
ds^2 = -(1+2\phi)d\tau^2 + a^2(\tau)(1-2\psi)\left(
d\tilde{r}^2 + \tilde{r}^2d\Omega^2\right)\,,
\label{nonlin-3eq1}
\end{equation}
with at least the conditions
\begin{equation}
\mid\phi\mid\ll1 ~~;~~ \mid\psi\mid\ll1\,,
\label{nonlin-3eq2}
\end{equation}
being satisfied.
Since $t$ is the proper
time of each matter shell, the quantity $\partial_t\tilde{r}$ is simply
the velocity of matter in the $(\tau,\tilde{r})$ frame (which is
comoving with the Hubble flow) :   
\begin{equation}
\tilde{v} \equiv \frac{\partial\tilde{r}}{\partial t}\,,
\label{nonlin-3eq8}
\end{equation}
is the radial comoving peculiar velocity of the matter shells in the 
$(\tau,\tilde{r})$ frame. 
We showed that the required transformation exists, {\it provided matter peculiar velocities
remain small}, which is consistent with what has een shown by other authors, and is true for the observed Universe.

In the cosmological equations derived from Macroscopic Gravity we already have in place the formalism for calculating the backreaction when the metric is of the perturbed FLRW form. From there it follows that the backreaction is very small, in the nonlinear structure formation regime, provided matter peculiar velocities are small. It can be argued that this result is independent of the assumption of spherical symmetry in the toy model. The situation could be very different though, if there are dominant nonlinear structures in today's Universe, comparable to the Hubble radius. 

\subsection{Perturbation theory around a background - the shortwave approximation} 

Green and Wald \cite{W2} have recently given an analysis of the growth of metric perturbations, assuming that the metric is always close to a given background, although matter perturbations can be arbitrarily large. No averaging of an underlying spacetime geometry is done, and it is assumed that there is a homogeneity length scale at around $100$ Mpc, much smaller than the Hubble radius.  It is shown that if the small-scale motions of matter inhomogeneities are non-relativistic, the deviations from the background metric are small, and well-described by Newtonian gravity. This result tallies with what has een found by others before, including us. It is further shown that subject to the matter satisfying weak energy condition, the effect of small scale inhomogeneities on large scale dynamics is to produce an effective trace-free stress energy tensor. One might ask if this traceless nature of the correction has to do with no averaging over finite volumes being carried out.

Thus the assumption of non-relativistic peculiar velocities along with the assumption of a homogeneity scale much smaller than the Hubble radius strongly suggest a negligible effect of small scale inhomogeneities on the average large-scale dynamics. The first of these two assumptions is well supported by observations. There is no observational evidence against the second assumption, but nor is it firmly established by observations. If this assumption is correct, either a small cosmological constant, or a modification of general relativity on large scales, is indicated by the observed cosmic acceleration. If this assumption turns out to be not correct, the effect of inhomogeneities could be significant, and remains an important question for further investigation.   

\vskip 0.4 in 

\noindent I would like to thank Aseem Paranjape for collaboration and extended discussions during the period 2006-2009 : our work on application of Macroscopic Gravity to cosmology would not have been possible without his ingenuity in simplifying the original system of equations. I would like to thank Friedrich Hehl for suggesting in the first place that we apply MG to cosmology. Correspondence and interactions with Roustam Zalaletdinov are gratefully acknowledged.  I am also thankful to Thomas Buchert for correspondence in the early stages of this work. It is a pleasure to thank the organizers of the conference for their kind hospitality, and the conference participants for stimulating discussions.

\bigskip

The list of references below is far from exhaustive, and references to a large number of the original papers on the subject can be found in the review articles cited here.

\end{document}